\documentclass[fleqn,10pt]{wlscirep}
\usepackage[utf8]{inputenc}
\usepackage[T1]{fontenc}

\usepackage{graphicx}% Include figure files
\usepackage{bm}% bold math
\usepackage{color}
\usepackage{enumerate}
\usepackage{xspace}
\usepackage{mathrsfs}
\newcommand{\magnetite}{\ensuremath{\mathrm{Fe}_3\mathrm{O}_{4}}\xspace}
\newcommand{\maghemite}{\ensuremath{\gamma-\mathrm{Fe}_2\mathrm{O}_{3}}\xspace}

\newcommand{\kc}{\ensuremath{k_{\mathrm{c}}}\xspace}
\newcommand{\kcc}{\ensuremath{\frac{\kc}{2}}\xspace}
\newcommand{\smsz}{\ensuremath{S_z}}
\newcommand{\smJijFeTT}{\ensuremath{J_{\mathrm{Fe}}^{\mathrm{TT}}}\xspace}
\newcommand{\smJijFeOO}{\ensuremath{J_{\mathrm{Fe}}^{\mathrm{OO}}}\xspace}
\newcommand{\smJijFeTO}{\ensuremath{J_{\mathrm{Fe}}^{\mathrm{TO}}}\xspace}
\newcommand{\smmuFeT}{\ensuremath{\mu_{\mathrm{Fe}}^{\mathrm{T}}}\xspace}
\newcommand{\smmuFeO}{\ensuremath{\mu_{\mathrm{Fe}}^{\mathrm{O}}}\xspace}
\newcommand{\smdip}{\ensuremath{\mathbf{B}_{\mathrm{dip}}}\xspace}
\newcommand{\smHapp}{\ensuremath{\mathbf{B}_{\mathrm{a}}}\xspace}
\newcommand{\sms}{\ensuremath{\mathbf{S}}\xspace}

\newcommand{\Si}{\ensuremath{\mathbf{S}_{\mathrm{i}}}\xspace}
\newcommand{\Heff}{\ensuremath{\mathbf{B}_{\mathrm{eff}}^{\mathrm{i}}}\xspace}

\newcommand{\vampire}{\textsc{vampire} }
\newcommand{\Tc}{\ensuremath{T_{\mathrm{c}}}\xspace}

\newcommand{\kB}{\ensuremath{k_{\mathrm{B}}}\xspace}
\newcommand{\muB}{\ensuremath{\mu_{\mathrm{B}}}\xspace}
\newcommand{\Ms}{\ensuremath{M_{\mathrm{s}}}\xspace}
\newcommand{\Kc}{\ensuremath{K_{\mathrm{c}}}\xspace}
\newcommand{\Keff}{\ensuremath{K_{\mathrm{eff}}}\xspace}
\newcommand{\Nd}{\ensuremath{N_{\mathrm{d}}}\xspace}
\newcommand{\Hd}{\ensuremath{\mathbf{H}_{\mathrm{d}}}\xspace}
\newcommand{\M}{\ensuremath{\mathbf{M}}\xspace}
\newcommand{\Hth}{\ensuremath{\mathbf{B}_{\mathrm{th}}}\xspace}

\newcommand*{\rttensor}[1]{\overline{\overline{#1}}}

\newcommand{\mcellvec}{\ensuremath{\vec{m}_{\mathrm{mc}}}\xspace}

\newcommand{\Vatom}{\ensuremath{V_{\mathrm{atom}}}\xspace}

\newcommand{\Bdipcellvec}{\ensuremath{\mathbf{B}_{\mathrm{dip,mc}}}\xspace}
\newcommand{\Bdipvec}{\ensuremath{\mathbf{B}_{\mathrm{dip,mc}}}\xspace}

\newcommand{\Diptens}{\ensuremath{\rttensor{\mathbf{D}}}\xspace}
\newcommand{\DemagFactorTensor}{\ensuremath{\rttensor{\mathbf{N}}}\xspace}

\title{The role of faceting and elongation on the magnetic anisotropy of magnetite Fe$_3$O$_4$ nanocrystals}

\author[1,2]{Roberto~Moreno}
\author[1]{Samuel~Poyser}
\author[1]{Daniel~Meilak}
\author[1]{Andrea~Meo}
\author[1]{Sarah~Jenkins}
\author[1]{Vlado~K.~Lazarov}
\author[1]{Gonzalo~Vallejo-Fernandez}
\author[3]{Sara~Majetich}
\author[1,*]{Richard~F.~L.~Evans}

\affil[1]{Department of Physics, The University of York, York, YO10 5DD, UK}
\affil[2]{Earth and Planetary Science, School of Geosciences, University of Edinburgh, Edinburgh EH9 3FE, UK}
\affil[3]{Physics Department, Carnegie Mellon University, Pittsburgh, Pennsylvania 15213, USA}
\affil[*]{richard.evans@york.ac.uk}

\begin{abstract}
\magnetite nanoparticles are one of the most promising candidates for biomedical applications such as magnetic hyperthermia and theranostics due to their bio-compatibility, structural stability and good magnetic properties. However, much is unknown about the nanoscale origins of the observed magnetic properties of particles due to the dominance of surface and finite size effects. Here we have developed an atomistic spin model of elongated magnetite nanocrystals to specifically address the role of faceting and elongation on the magnetic shape anisotropy. We find that for faceted particles simple analytical formulae overestimate the magnetic shape anisotropy and that the underlying cubic anisotropy makes a significant contribution to the energy barrier for moderately elongated particles. Our results enable a better estimation of the effective magnetic anisotropy of highly crystalline magnetite nanoparticles and is a step towards quantitative prediction of the heating effects of magnetic nanoparticles.
\end{abstract}
\begin{document}

\flushbottom
\maketitle

\section*{Introduction}
Magnetic fluid hyperthermia \cite{Pankhurst,Tartaj2003,Dutz_2014,PerigoAPR2015} is a promising treatment for brain and prostate cancers due to the localized nature of the treatment compared to chemo or radiotherapy. Brain cancers in particular are difficult to treat with conventional therapies due to the sensitivity of the surrounding tissue with typically less than 20\% survival rate after 10 years. Magnetic nanoparticles used in magnetic hyperthermia must be bio-compatible and provide efficient and reliable heating. Iron oxide nanoparticles in the form of magnetite (\magnetite) and maghemite (\maghemite) are often promoted as the only viable candidate materials for hyperthermia due to the combination of resistance to oxidation, high saturation magnetization and moderate magnetic anisotropy~\cite{LAURENT2011}. When coated with suitable surfactant molecules these particles are highly bio-compatible~\cite{LAURENT2011} and can be used for magnetic hyperthermia, targeted drug delivery~\cite{Kumar} and contrast enhancement for Magnetic Resonance Imaging (MRI) and Ultrasound (US) imaging. 

In practical applications both the effective magnetic anisotropy of the nanoparticles $\Keff$ and the particle volume $V$ determine the magnetic thermal stability with energy barrier $\Delta E = \Keff V$. It is important to note that the effective anisotropy is much smaller for cubic anisotropy than would be expected from the usual anisotropy constants $K_1$ and $K_2$ due to the nature of the energy surface. Specially the effective anisotropy is $\Keff = K_1/4$ for [100] easy and $\Keff = K_1/12$ for [111] easy axis systems. For optimal heating properties the particle size is often tuned, but depending on the particle size, morphology and preparation conditions the effective anisotropy can be significantly different from the bulk cubic anisotropy of magnetite. In the literature this effect is often attributed to the presence of surface anisotropy~\cite{BodkerPRL1994,evansMRS2013} arising at the surface of the nanoparticle due to changes in atomic symmetry at the particle surface. Additional anisotropy effects may come bulk~\cite{Nedelkoski2017} and surface defects such as spin canting resulting from Dzyaloshinskii–Moriya interactions~\cite{Oberdick2018} and also through magnetic dipole-dipole interactions due to particle agglomeration~\cite{Burrows2010,HaasePRB}. In most of these studies however the effects of particle shape and elongation have been largely ignored due to the computational complexity. Recent experimental measurements of particle size distributions~\cite{Vallejo-FernandezAPL2013} have suggested that shape anisotropy is likely to play an important role in the magnetic anisotropy of magnetite samples, particularly at small sizes, and is also a feature of crystals in magnetotactic bacteria~\cite{ChariaouBio2015}. 

Surface anisotropy in metallic thin films is often associated with heavy metal elements such as Pt, Pd and Ir with large spin-orbit coupling. For surfactant-coated magnetic oxide nanoparticles the surface termination is with organic molecules with low atomic mass and consequently low spin-orbit coupling effects. When considering a simple surface ($K_{\mathrm{S}}$) and volume ($K_{\mathrm{V}}$) anisotropy contributions, the effective anisotropy can be expressed as 
\begin{equation}
    \Keff = K_{\mathrm{V}} + \frac{6  K_{\mathrm{S}}}{d}
\end{equation}
where $d$ is the particle diameter. In reality the effects of surface anisotropy are much more complex~\cite{JametPRL,JametPRB2004,Salazar-Alvarez2008,AlliaJAP2014}, particularly in the case of large surface anisotropy which can lead to significant distortions of the atomic spin structure~\cite{GaraninPRL,KACHKACHIJMMM2000,GaraninPRB2018,YanesPRB2007,LabayeJAP2002} and unusual temperature dependent effects \cite{YanesJPhysD2010}. For oxide nanoparticles extensive simulations using transverse and N\'eel surface anisotropy models have been used to explain the observed increase of magnetic anisotropy in nanoparticles~\cite{KACHKACHIJMMM2000,Martinez-Boubeta2013,IglesiasPRB2001,RESTREPO2006221,Kachkachi2000,KodamaPRB1999,MAZOZULUAGA2007187,ZuluagaJAP2008}.  In the case of a cleaved (001) magnetite surface the intrinsic surface anisotropy is very large~\cite{CoeyFe3O4JAP1993} yet this is unlikely to be structurally stable due to the polar nature of the interface~\cite{ParkinsonSSR2016}, and surface passivation by oxygen removes the effective surface anisotropy contribution~\cite{CoeyFe3O4JAP1993}. Conversely passivation with C or H adatoms may still retain a strong magnetic surface anisotropy while stabilizing the structure, where Fe-O hybridization will induce strong surface anisotropy effects.

The relative balance of surface anisotropy, shape anisotropy and interparticle interactions contributions to the effective anisotropy is still an open question. In particular the specific contribution of the magnetic shape anisotropy for small nanoparticles of magnetite with faceted and unfaceted surfaces is only approximated by simple analytical treatments assuming ellipsoidal shapes, limiting the ability to accurately determine the effects in realistic nanoparticle systems. Furthermore standard numerical treatments often assume a uniaxial bulk anisotropy for simplicity, whereas the bulk anisotropy has an underlying cubic symmetry. This is especially important when considering the energy barrier which is a factor \Kc/4 smaller for positive cubic anisotropy and \Kc/12 smaller for negative cubic anisotropy as seen in magnetite and maghemite.

In this paper we present an atomistic spin model of \magnetite explicitly including atomic level anisotropies and Heisenberg exchange interactions. Using a simple algorithm to generate elongated and faceted nanocrystals we apply our model to determine the shape anisotropy, effective anisotropy and energy barriers of magnetite nanocrystals of different shapes and sizes. We find that the underlying cubic magnetic anisotropy has an important effect for moderately elongated ($<20\%$) particles where the magnetization has a shallow energy minimum at low angles from the elongation axis characteristic of a higher order uniaxial magnetic anisotropy. This effect is particularly important for numerical studies of particle interaction effects where the assumed magnetic anisotropy is often exaggerated.

%----------------------------------------------------------------------------------------------------------------------------------
% Results
%----------------------------------------------------------------------------------------------------------------------------------
\section*{Results}

\subsection*{Modelling magnetite nanocrystals}

\begin{figure}[!t]
\center
\includegraphics[width=7.5cm, trim=0 0 0 0]{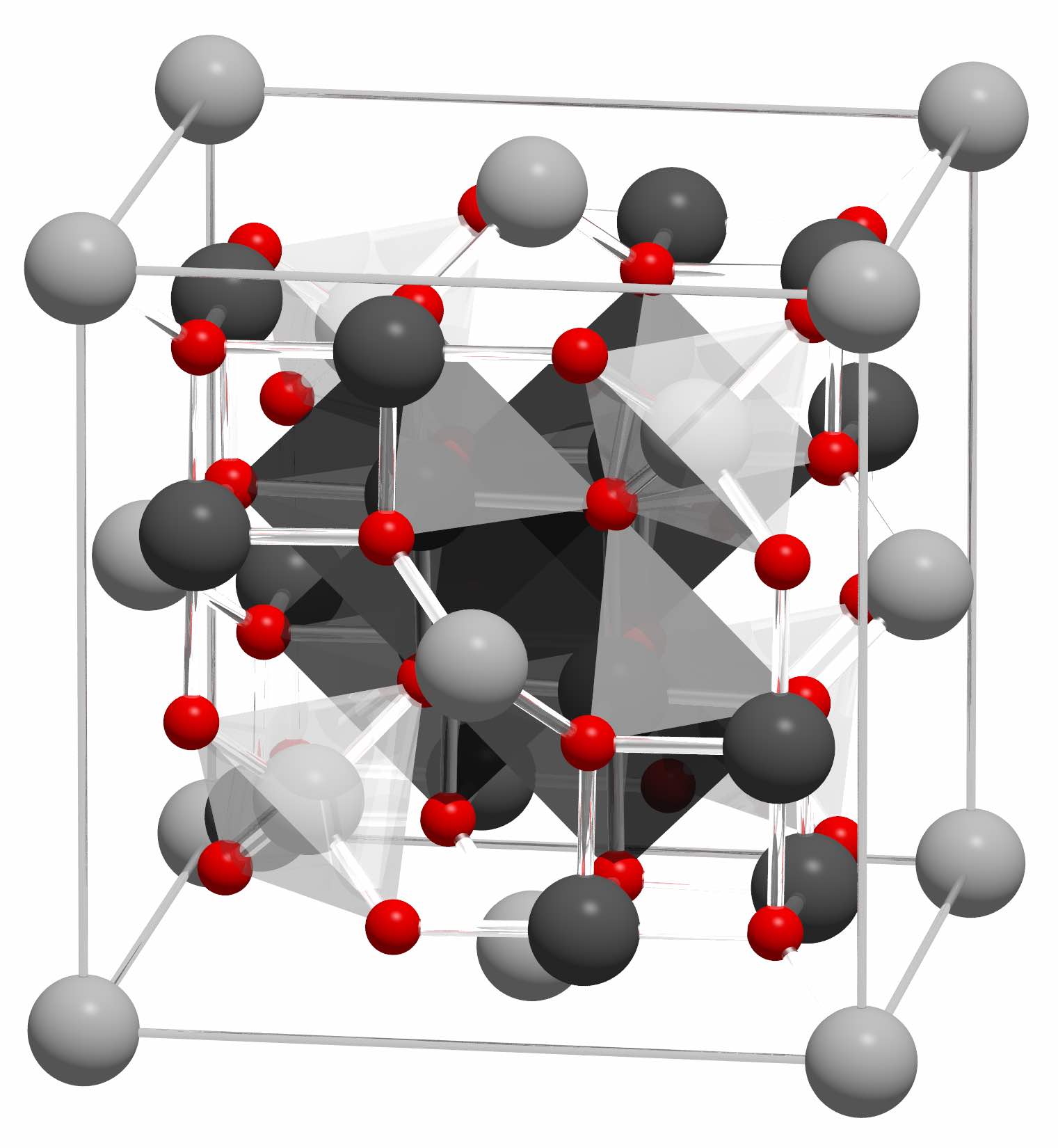}
\caption{Visualization of the magnetite (\magnetite) unit cell identifying Octahedral Fe $^{2.5+}$ (dark grey), Tetrahedral Fe $^{2+}$ (light grey) and Oxygen (red) sites. The local site symmetries are shown by octahedra and tetrahedra around fully coordinated Fe sites within the unit cell. The different bond angles between Fe sites leads to a dominant antiferromagnetic coupling between FeT and FeO sites, giving a bulk ferrimagnetic order. Magnetite has an inverse spinel structure with a lattice constant of 8.397\r{A}.}
\label{fig:unitcell}
\end{figure}

To model the effects of particle size, faceting and elongation on the effective magnetic anisotropy we construct an atomistic spin model including details of the crystal structure of magnetite within the simulations. The magnetite unit cell \cite{ShullPhysRev1951,Gorter1955} consists of 24 magnetic atoms with 8 tetrahedrally coordinated $(\mathrm{Fe}^{3+})$ sites and 16 octrahedrally coordinated mixed valence $(\mathrm{Fe}^{2.5+})$ sites shown schematically in Fig.~\ref{fig:unitcell}. The oxygen atoms are assumed to be non-magnetic but mediate super-exchange interactions between the Fe sites.

\begin{figure*}[!htb]
\center
\includegraphics[width=13.6cm, trim=0 0 0 0]{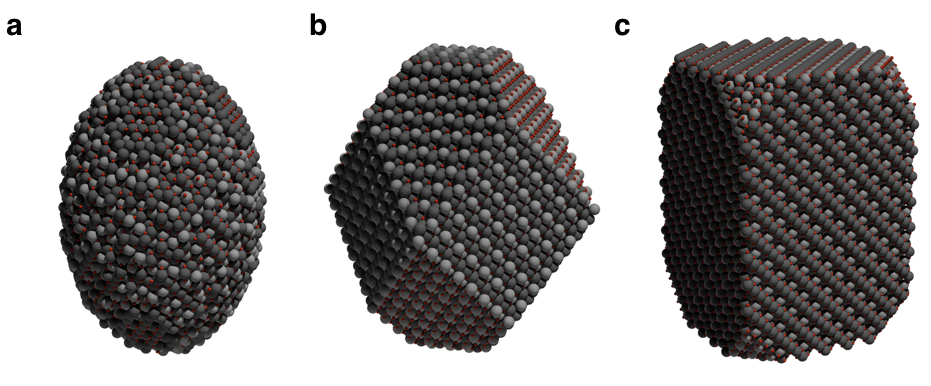}
\caption{Schematic illustration of generated elongated particles with size of 6 nm and elongation of 50\% with ellipsoidal (a), truncated octahedral (b) and cuboidal (c) shape. Spheres indicate octahedral iron atoms (dark grey), tetrahedral iron atoms (light grey) and oxygen atoms (small, red).}
\label{fig:particles}
\end{figure*}

Magnetite nanoparticles can form both regular and irregular shapes, sometimes with and without clear surface faceting depending on how samples are prepared experimentally. Defects such as anti-phase boundaries are also commonly observed in magnetite particles \cite{Nedelkoski2017}, leading to a wide range of magnetic properties with complex size and shape dependencies. For medical applications such as magnetic hyperthermia \cite{Pankhurst,Kumar} the magnetic properties of the particles need to be precisely controlled, essentially requiring highly ordered single-crystal particles within a narrow size range. For such highly crystalline nanoparticles the surface morphology is typically spherical where surfactants are used in the preparation (by passivating the surface) or otherwise highly faceted due to surface charge ordering, since polar surfaces are energetically unfavorable \cite{ParkinsonSSR2016}. From structural studies~\cite{Faivre,JayaNanoscale2015,ZhouCoM2015} it has been found that (111), (110) and (100) facets are the most commonly observed in synthetic magnetite nanoparticles. We have implemented a simple geometric method to recreate realistically shaped nanoparticles with arbitrary facets and elongation as described in the methods section. In the present paper we consider three kinds of particles: ellipsoids, truncated octahedra and cuboids, each shown schematically in Fig.~\ref{fig:particles}.

\subsection*{Calculation of the demagnetizing factor}
A uniformly magnetized nanoparticle has a magnetic field which directly opposes the magnetization of the particle, known as the demagnetizing field \cite{Jiles}. The demagnetizing field $\Hd$ is expressed as a tensor-vector product with the magnetization given by
\begin{equation}
\Hd = -\DemagFactorTensor \cdot \M,
\label{eq:Hdip_mean_field}
\end{equation}
where $\M = (\mathbf{m}\cdot\Ms)$ is the vector magnetization, \Ms is the saturation magnetization, $\mathbf{m}$ is a unit vector describing the direction of the magnetization and \DemagFactorTensor is the demagnetization tensor that in Cartesian coordinates can be expressed as:
\begin{align}
\label{eq:demagnetization_tensor}
\DemagFactorTensor = 
\left(
	\begin{array}{ccc} 
       N_{xx}  & N_{xy}  & N_{xz} \\
       N_{yx}  & N_{yy}  & N_{yz} \\
       N_{zx}  & N_{zy}  & N_{zz}
	\end{array}
\right).
\end{align}
The demagnetization tensor \DemagFactorTensor depends on the shape and geometry of the sample and for general systems with inhomogeneous magnetization or complex shapes analytic expressions do not exist. For simple elongated systems we study here the magnetostatic field inside the body is uniform and the off-diagonal components of \DemagFactorTensor are close to zero. In this case one can express the full tensor as a trace with eigenvalues $N_{xx},N_{yy},N_{zz}$
\begin{align}
\label{eq:demagnetization_trace}
\mathrm{tr}\left(\DemagFactorTensor\right) = 
\left(
	\begin{array}{ccc} 
       N_{xx}  & 0       & 0 \\
       0      & N_{yy}  & 0 \\
       0      & 0       & N_{zz}
	\end{array}
\right)
\end{align}
which is usually expressed as a demagnetizing factor $\Nd = \mathrm{tr}\left(\DemagFactorTensor\right)$ and so the demagnetizing field reduces to 
\begin{equation}
	\Hd = -\Nd \M \mathrm{.}
\end{equation}
 The demagnetizing factor can be derived analytically for certain geometric shapes such as ellipsoids \cite{Osborn} and rectangular prisms \cite{Aharoni}. For realistically shaped particles with complex facets and crystal structures the simple analytical formulae do not apply due to deviations from a regular shape and discontinuous nature of the magnetization formed from localized atomic magnetic moments. To calculate the effective demagnetizing field we discretize our system into (1 nm)$^3$ cells and compute the full demagnetizing tensor for each cell according to Eq.~\ref{eq:dip} and then average the contributions of each cell to give the total demagnetizing tensor for the system.

\begin{figure}[!tb]
\center
\includegraphics[width=12.0cm, trim=20 0 0 30]{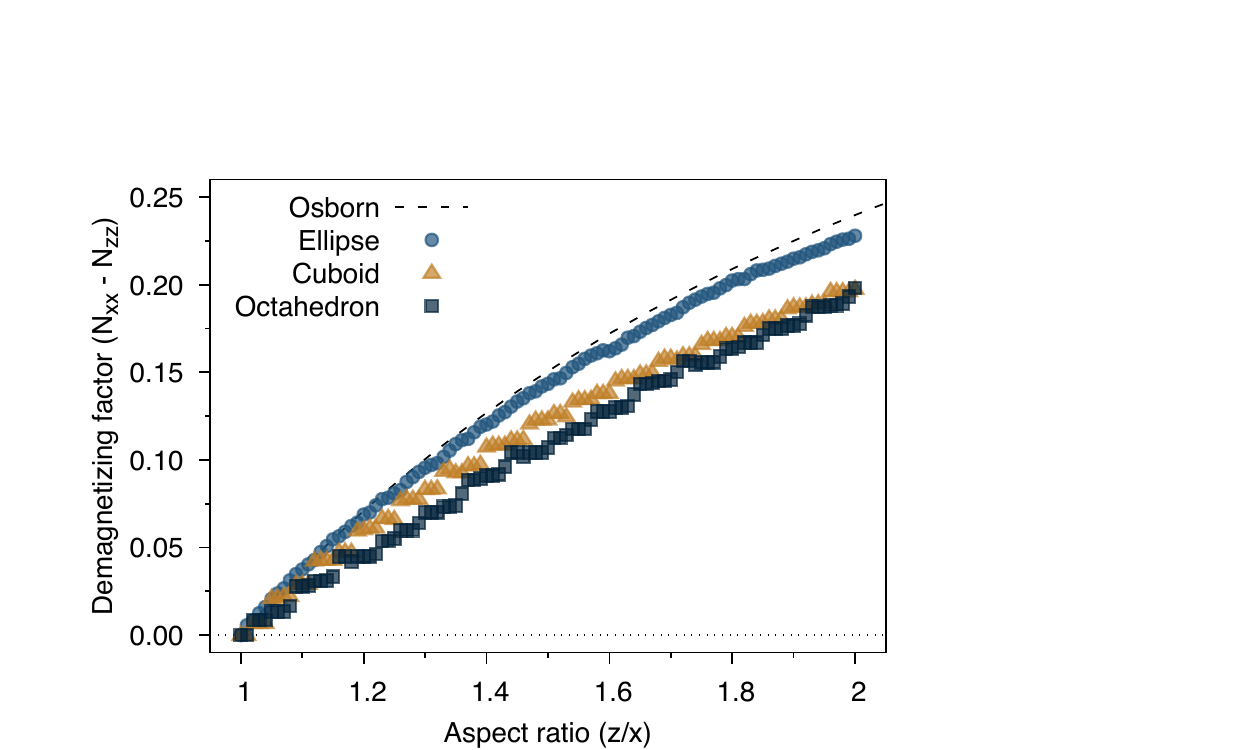}
\caption{Plot of the elongation dependence of the demagnetizing factor $N_{xx}-N_{zz}$ for different particle shapes 6 nm in size. The data show a small deviation from the Osborn relation for a uniformly magnetized ellipsoid at larger elongations likely due to the small number of atoms at the outer extremities of the particle. The faceted cuboid and truncated octahedron nanoparticles show a similar qualitative behaviour with discrete steps in the effective demagnetizing factor and an asymptotically lower demagnetizing factor than for a regular ellipsoid.}
\label{fig:shape}
\end{figure}

\subsection*{Shape and size dependence of the demagnetizing factor}
We first consider the role of particle faceting and elongation on the demagnetizing factor of magnetite nanocrystals. We assume that the magnetization of the particle is formed of a uniform ferrimagnetic ground state with perfect alignment/anti-alignment of spins with the $z$-axis, which is a reasonable approximation at temperatures much lower than the Curie temperature. We compute the demagnetizing tensor for each of the particles using the inter-intra macrocell method with a macrocell size of 1 nm. For the facets as shown in Fig.~\ref{fig:particles} we compare the calculated demagnetizing factor $(N_{xx} - N_{zz})$ for a fixed lateral dimension of 6 nm diameter (along the $x,y$ directions) and different elongations along the $z$-axis, leading to an increase in the shape anisotropy. The calculated demagnetizing factor as a function of aspect ratio is shown in Fig.~\ref{fig:shape}. We also compare our results with the analytical formula of Osborn \cite{Osborn} for the demagnetizing factor of a regular ellipsoid as function of the aspect ratio $k_0 = z/x$
\begin{align}
\label{eq:osborn_demag}
	\begin{array}{lcc}
      N_{zz} = \frac{1}{1-k_0^2} \left[ 
      1 - \frac{k_0}{\sqrt{1-k_0^2}}\arccos\left(k_0\right)
      \right] &  & \text{for }k_0<1 \\
     N_{zz} = \frac{1}{3}  &  & \text{for }k_0=1 \\
      N_{zz} = \frac{1}{k_0^2-1} \left[ 
      \frac{k_0}{\sqrt{k_0^2-1}}\operatorname{arcosh} \left(k_0\right) -1
      \right] &  & \text{for }k_0>1 \mathrm{.}
	\end{array}
\end{align}
where the other two diagonal components of \DemagFactorTensor can be obtained by the symmetry relation $N_{xx} + N_{yy} + N_{zz}=1$ in SI units. At zero elongation (aspect ratio = 1) the effective demagnetizing factor is zero for all particles, as expected. With increasing elongation the demagnetizing factor increases for all particle shapes. In the case of the ellipsoid particle shape small elongations closely follow Osborn's formula for a regular ellipsoid, but deviate slightly for large aspect ratios. This is due to the relatively large unit cell of magnetite which leads to significant deviations from a perfect ellipsoidal shape at the extrema along the $z$-axis. For the cuboidal and truncated octahedral particles the demagnetizing factor is systematically lower than that of an ellipsoid with increasing aspect ratio. In itself this is not surprising since different particle shapes will give different demagnetizing fields. However, it is important to note that a general assumption that Osborn's formula for the aspect ratio dependence of the demagnetizing factor is correct for faceted particles is a bad approximation, and will lead to an overestimation of the effective magnetic anisotropy in nanoparticle systems.

\begin{figure}[!tb]
\center
\includegraphics[width=12cm, trim=20 0 0 30]{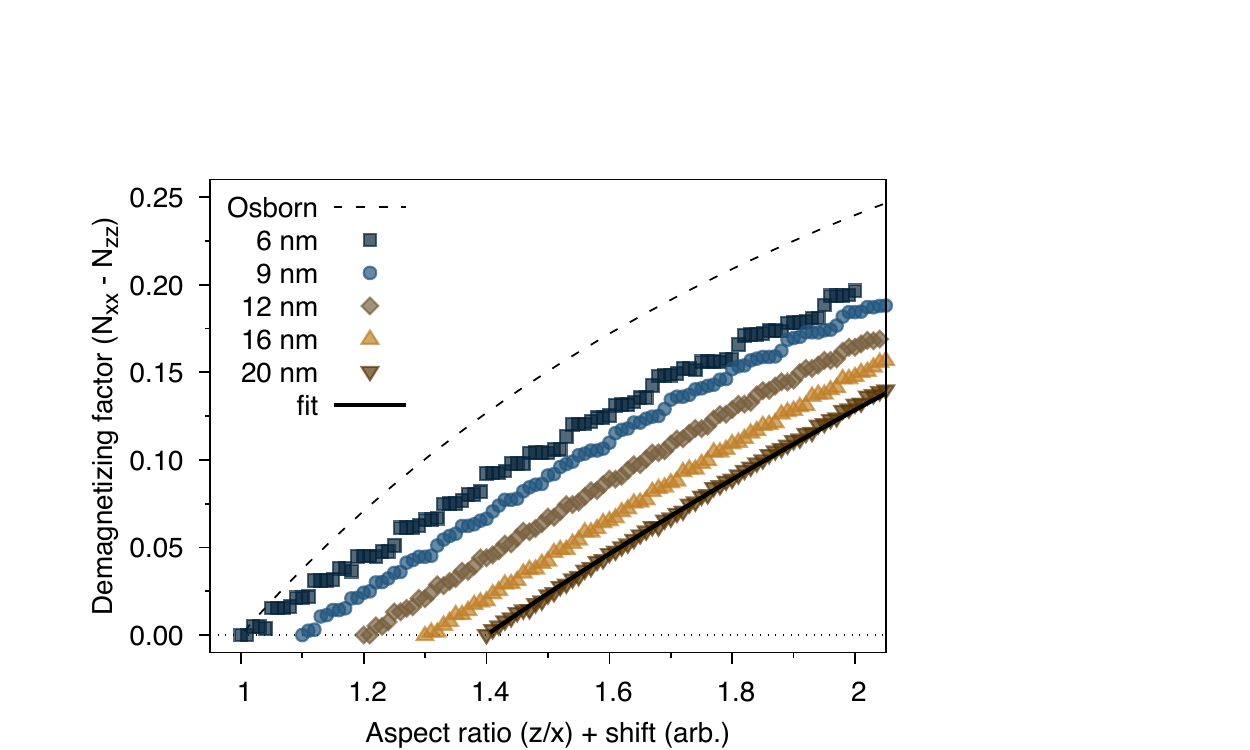}
\caption{Plot of the aspect ratio dependence of the demagnetizing factor for truncated octahedra as a function of particle size. An arbitrary shift of 0.1 is applied to the aspect ratio to show the data for different sizes since their envelope essentially overlaps. As the particle size increases the discontinuous nature of the demagnetizing factor with increasing elongation reduces. In the limit of large particle sizes the demagnetizing factor is significantly different from Osborn's relation for an ellipsoid and fitted by Eq.~\ref{eq:dm_oct}.}
\label{fig:size}
\end{figure}

An interesting effect for faceted particles is the discontinuous nature of the demagnetizing factor with increasing aspect ratio. This is due to the nanoscale size of the system and so incremental increases in the aspect ratio will not always be sufficient to add a new layer of atoms on a particular facet. This effect is particular to the type of faceting and particle size and in general difficult to predict without a detailed atomistic model of the particles. To investigate the asymptotic limit of the demagnetizing factor for the truncated octahedral particles we show the aspect ratio dependence of the demagnetizing factor for different lateral particle sizes in Fig.~\ref{fig:size}.

As the particle size increases the discontinuous changes in the demagnetizing factor reduce as expected, as the particle shape is better able to follow the exact aspect ratio. In all cases the demagnetizing factor is significantly lower than that of Osborn's formula for an ellipsoid for all aspect ratios. In the asymptotic limit for the largest truncated octahedral particle size, we can fit the aspect ratio dependence of the demagnetizing factor up to aspect ratios of $z/x \leq 2$ with the formula

\begin{equation}
N_{xx} - N_{zz} = A \left(\sqrt{\frac{z}{x}}-1\right)
\label{eq:dm_oct}
\end{equation}

\noindent where $A = 0.4861 \pm 0.0007$. Eq.~\ref{eq:dm_oct} provides a better estimation of the aspect ratio dependence of the demagnetizing factor in the case of faceted particles compared to the formula for a regular ellipsoid which is important for experimental estimations of the effective magnetic anisotropy \cite{Vallejo-FernandezAPL2013} and also for theoretical and numerical calculations of the heating properties of particulate systems \cite{Burrows2010,Ruta2015,TanPRB2014}.

\subsection*{Shape and elongation dependence of the effective magnetic anisotropy energy}

\begin{figure*}[!tb]
\center
\includegraphics[width=17.3cm, trim=0 0 0 0]{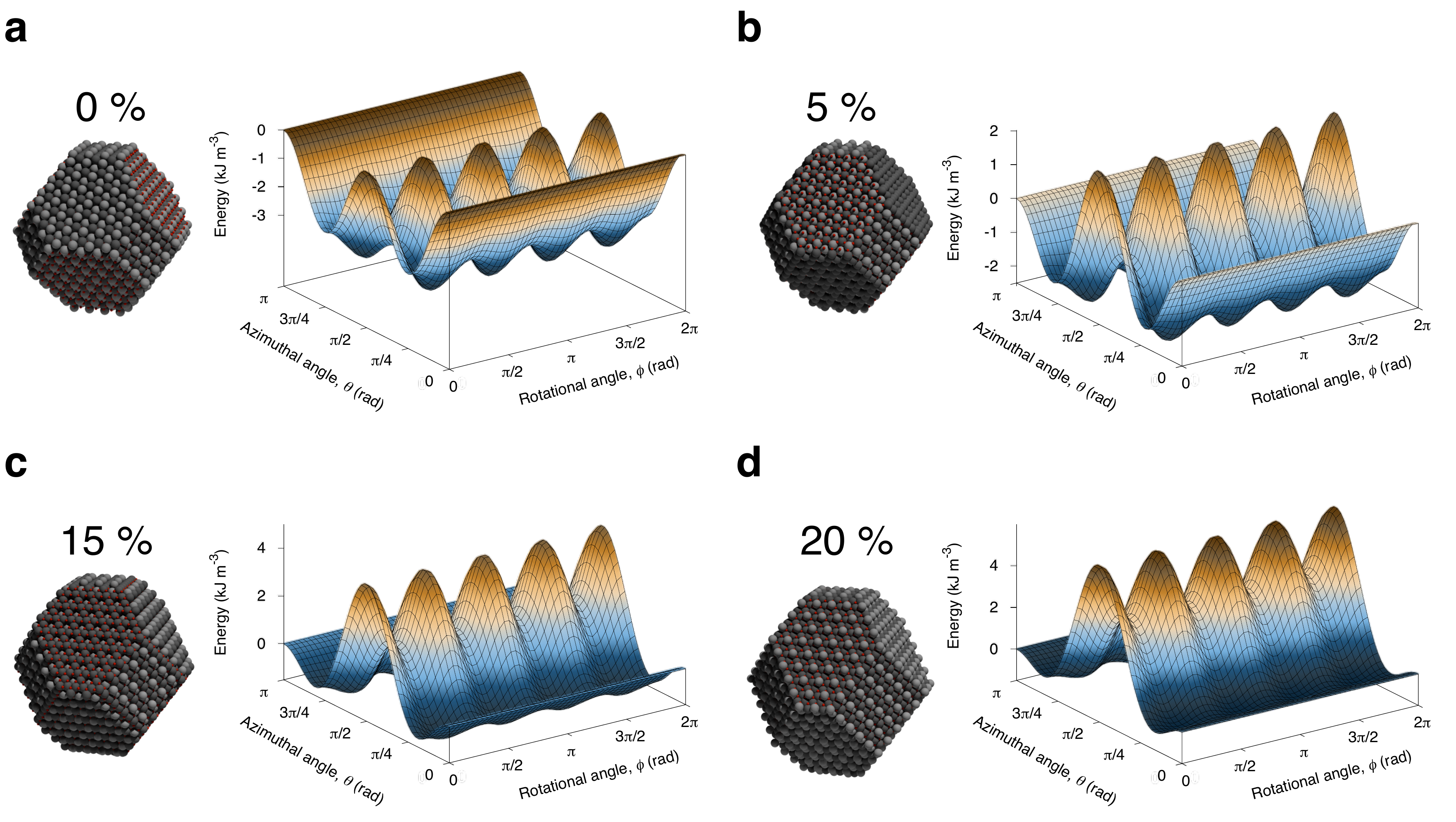}
\caption{Plots of the effective anisotropic energy surface for elongated truncated octahedral nanoparticles with lateral size of 6 nm along the $(x,y)$ directions and elongation 0\% (a), 5\% (b), 15\% (c), and 20\% (d). The data show that the cubic anisotropy makes a significant contribution to the effective anisotropy for all elongations with rotational energy barriers existing up to 15\% elongation.}
\label{fig:surface}
\end{figure*}

The weak magnetocrystalline anisotropy of magnetite means that shape effects are expected to dominate the magnetic anisotropy of nanoparticle systems~\cite{Vallejo-FernandezAPL2013}. In numerical calculations the effective anisotropy energy of the nanoparticles is often assumed to be completely uniaxial~\cite{Burrows2010,Ruta2015,TanPRB2014,HaasePRB,Serantes2010} and of the form 

\begin{equation}
E_{\mathrm{k}} = -K_{\mathrm{eff}} \left(\mathbf{m} \cdot \mathbf{e} \right)^2 \equiv +\Keff \sin^2 \theta
\label{eq:keff}
\end{equation}

\noindent where $K_{\mathrm{eff}}$ is a volumetric effective anisotropy energy, $\mathbf{m}$ is a unit vector describing the direction of the magnetization, $\mathbf{e}$ is a unit vector describing the easy axis of the magnetization or equivalently $\theta$ is the angle of the magnetization from the easy axis $\mathbf{e}$. With our detailed calculation of the size and aspect ratio dependence we are now able to test this assumption. The shape anisotropy arises from the anisotropic contribution of the demagnetizing energy, where the demagnetizing energy $E_{\mathrm{d}}$ is given by
\begin{equation}
E_{\mathrm{d}} = \frac{1}{2}\mu_0 \M \cdot \mathbf{H}_d
\label{eq:demag_energy}
\end{equation}
and $\mu_0 \equiv 4\pi \times 10^{-7}$ is the permeability of free space. In the case of an elongated particle with a symmetric trace demagnetizing factor expanding the vector-matrix-vector product 
\begin{equation}
E_{\mathrm{d}} = \frac{\mu_0}{2}
\left(
	\begin{array}{ccc} 
	M_x  & M_y  & M_z
	\end{array}
\right)
\left(
	\begin{array}{ccc} 
       N_{xx}  & 0       & 0 \\
       0      & N_{yy}  & 0 \\
       0      & 0       & N_{zz}
	\end{array}
\right)
\left(
	\begin{array}{c} 
	M_x \\
	M_y \\
	M_z
	\end{array}
\right)
\end{equation}
gives 
\begin{equation}
    E_{\mathrm{d}} = \frac{\mu_0}{2} \left(N_{xx} M_x^2 + N_{yy} M_y^2 + N_{zz} M_z^2\right)\mathrm{.}
\end{equation}
For a uniformly magnetized particle elongated along the $z$-axis the perpendicular components of the demagnetizing energy are the same. Considering the change in energy for a rotation of the magnetization in the $x-z$ plane the $z$ and $x$ components of the magnetization are related by $M_x^2 = (1-M_z^2)$. Applying the substitution $M_z = m_z \Ms$ yields an anisotropic contribution to the demagnetizing energy of
\begin{equation}
    E_{\mathrm{d}} = \frac{\mu_0 \Ms^2}{2} \left(N_{zz} - N_{xx}\right) m_z^2 + \mathrm{const.}
\end{equation}
which has the form of an effective shape anisotropy with constant
\begin{equation}
    K_{\mathrm{shape}} = -\frac{\mu_0 \Ms^2}{2} \left(N_{zz} - N_{xx}\right) = +\frac{\mu_0 \Ms^2}{2} \left(N_{xx} - N_{zz}\right)
\end{equation}
noting the change in sign. For a particle elongated along the $z$-axis $N_{xx} > N_{zz}$ and so the total shape anisotropy is positive. The total volumetric effective magnetic anisotropy can therefore be described as a summation of the intrinsic magnetocrystalline anisotropy and the effective demagnetizing energy given by:

\begin{equation}
	K_{\mathrm{eff}}(\mathbf{m}) = +\frac{\Kc}{2} \left(m_x^4 + m_y^4 + m_z^4\right) + \frac{\mu_0 \Ms^2}{2} \left(N_{xx} - N_{zz}\right)m_z^2
	\label{eq:effK}
\end{equation}
where $\mathbf{m}$ is a unit vector with components $m_x, m_y, m_z$ describing the direction of the magnetization of the particle, \Kc is the effective volumetric cubic anisotropy and \Ms is the saturation magnetization. The zero-temperature parameters \Kc and \Ms are both calculated from the atomistic model of the particle by taking summations over the atoms in the particle:
\begin{equation}
\Kc = \frac{\sum_{l=1}^n \kc}{n \Vatom}
\end{equation}
and
\begin{equation}
\Ms = \frac{\sum_{l=1}^n \mu_l \smsz}{n \Vatom}
\end{equation}
where $n$ is the number of magnetic atoms in the particle, $l$ is the atom number, \kc is the atomistic cubic anisotropy constant, $\mu_l$ is the local atomic moment of atom $l$, \smsz is the z-component of the local spin assuming all spins are aligned either parallel (octahedral sites, $\smsz = +1$) or antiparallel (tetrahedral sites, $\smsz = -1$) to the $z$-axis, and $\Vatom = a^3/24$ is the effective \textit{magnetic} volume of each atom, where $a=8.68$ \AA \xspace is the length of the magnetite unit cell. Considering applications to magnetic hyperthermia we use effective room-temperature parameters for the saturation magnetization $\Ms$(300 K)$ = 4.6 \times 10^5$ JT$^{-1}$m$^{-3}$ (equivalent to $4.6 \times 10^5$ A/m) and cubic anisotropy constant \Kc(300 K) = -11.5 kJ/m$^3$ where the negative sign indicates a preference for the (111) crystal directions. The prefactor in the effective shape anisotropy $\mu_0 \Ms^2 / 2 \sim 133$ kJ/m$^3$ is an order of magnitude larger than the magnetocrystalline anisotropy and so shape effects are expected to dominate the overall magnetic anisotropy of magnetite nanoparticles.

In Fig.~\ref{fig:surface} we plot the orientation dependence in spherical polar coordinates $(\theta, \phi)$ of the effective anisotropy energy for a 6 nm lateral ($x,y$) particle size for different elongations in the range 0-20\% typically seen for uniform experimental samples \cite{Vallejo-FernandezAPL2013} where elongated particles are not deliberately engineered \cite{ROCA2018}. For zero elongation Fig.~\ref{fig:surface}(a) the energy surface is completely cubic, with 8 energy minima corresponding to the [111] crystal directions of magnetite. For slightly elongated particles Fig.~\ref{fig:surface}(b) the demagnetizing energy adds a uniaxial component to the anisotropy energy lifting the maxima around the $x,y$-plane and reducing the rotational energy barrier between the cubic minima around the $\pm z$-axes. With increasing elongation Fig.~\ref{fig:surface}(c) the total effective anisotropy becomes more uniaxial, but a cubic contribution is still evident along the maxima and in the low-angle dependence of the effective anisotropy energy. The effects of the cubic anisotropy still persist even for 20\% elongation where the shape anisotropy contribution is five times the magnetocrystalline anisotropy.

\begin{figure}[!tb]
\center
\includegraphics[width=8.5cm, trim=0 0 0 0]{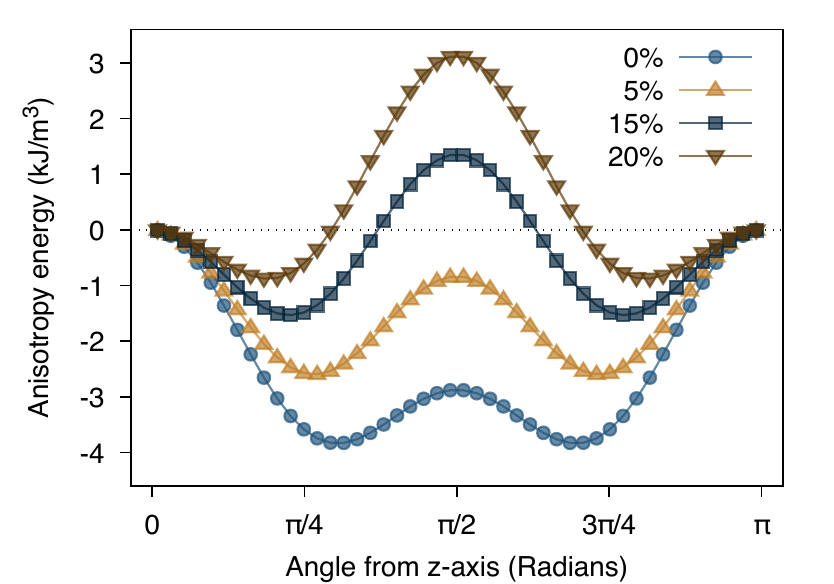}
\caption{Plot of the angle dependence of the effective anisotropy energy along the elongation axis $(\phi = \pi/4)$  comparing the full cubic anisotropy (lines) given by Eq.~\ref{eq:effK} with the higher order uniaxial approximation (points) given by Eq.~\ref{eq:effKU} for different particle elongations.  }
\label{fig:approx}
\end{figure}

In simple analytical and numerical models~\cite{Burrows2010,Ruta2015,HaasePRB} of magnetic hyperthermia the nanoparticles are assumed to have a simple uniaxial anisotropy arising from the difficulty of treating multiple energy minima associated with the cubic anisotropy. In the case of magnetite where shape effects dominate, it is possible to approximate the anisotropy as a higher order uniaxial magnetic anisotropy considering the minimum energy path along the elongation axis and given by
\begin{equation}
	K_{\mathrm{eff}}(m_z) = +\frac{\Kc}{2} \left(\frac{3}{2}m_z^4 - m_z^2\right) + \frac{\mu_0 \Ms^2}{2} \left(N_{xx} - N_{zz}\right)m_z^2
	\label{eq:effKU}
\end{equation}
neglecting arbitrary constants. Angle dependent plots of the effective uniaxial energy surface approximating the cubic energy surface are shown in Fig.~\ref{fig:approx} for different particle elongations. Except in the case of a perfectly symmetric particle where rotational and azimuthal energy barriers are the same,  Eq.~\ref{eq:effKU} describes the minimum energy path along the elongation axis of the particle with two unique energy minima. In this case the cubic contribution to the effective anisotropy energy is clearly visible, with minima far from the $\theta = 0,\pi$ minima expected for a purely second order uniaxial anisotropy term for moderately elongated particles. In the case of larger elongations $>20$\% the energy minima are shallow near the $\theta = 0,\pi$ directions allowing for large excursions of the magnetization from the poles. The higher order uniaxial approximation gives a better representation of the symmetry of the effective anisotropy energy for elongated particles while retaining two energy minima maintaining simplicity for analytical and numerical models.

\subsection{Shape and elongation dependence of the energy barrier}
\begin{figure}[!htb]
\center
\includegraphics[width=8.6cm, trim=0 0 0 0]{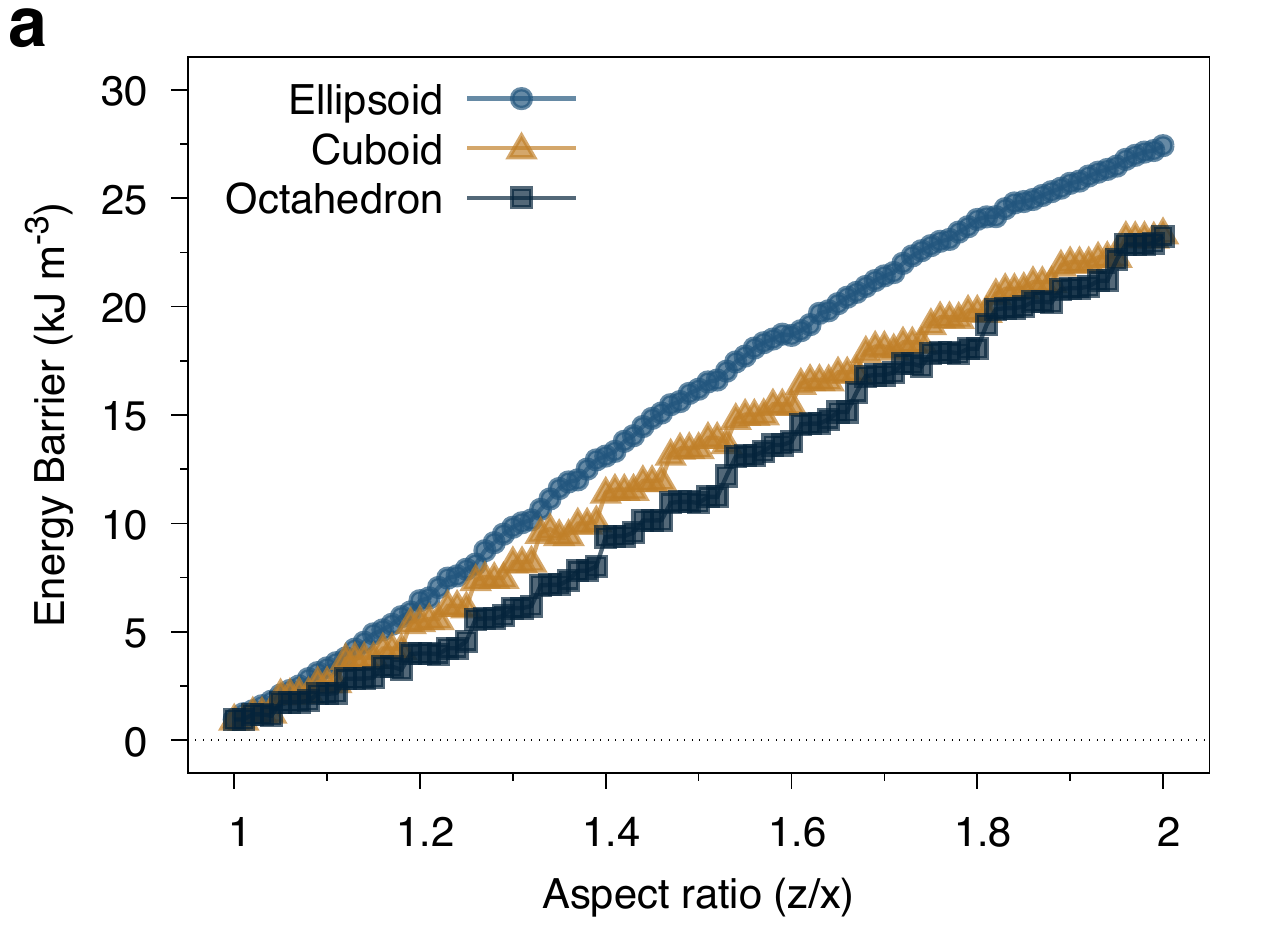}
\includegraphics[width=8.6cm, trim=0 0 0 0]{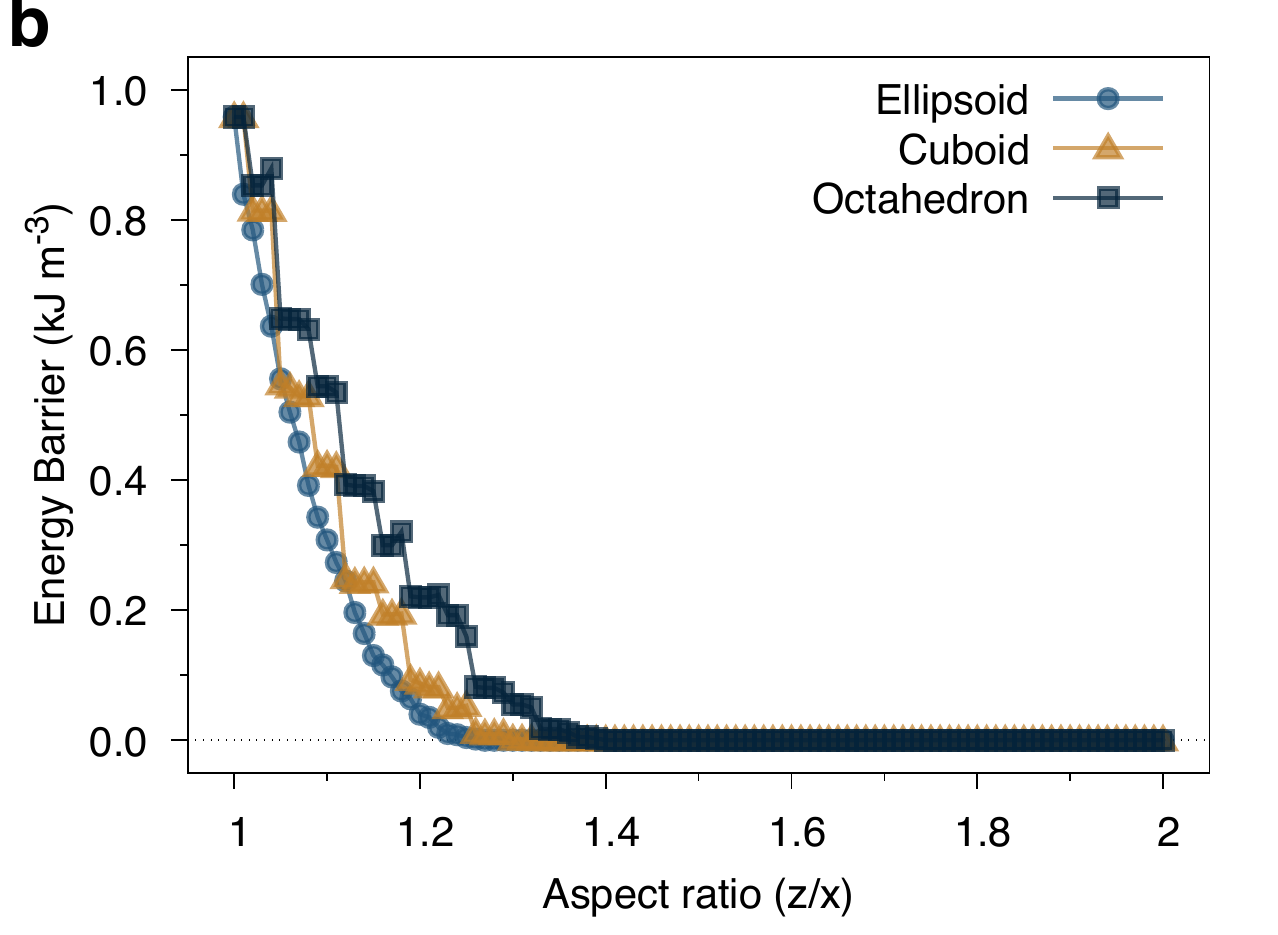}
\caption{Plot of the aspect ratio  dependence of the azimuthal (a) and rotational (b) energy barriers for different particle shapes and 6 nm particle size. For small elongations the azimuthal energy barrier rapidly increases and dominates over the rotational barrier arising from the cubic anisotropy. The rotational (cubic) character of the energy barrier along the energy minimum disappears for an elongation of 40\% where only a uniaxial barrier remains.}
\label{fig:eb_shape}
\end{figure}

The magnetic relaxation of superparamagnetic nanoparticles is determined by the energy barrier $\Delta E$ between stable minima and governed by the Arrhenius-N\'eel law

\begin{equation}
	\tau = \frac{1}{f_0} \exp \left(\frac{\Delta E}{\kB T}\right)
	\label{eq:an}
\end{equation}
where $\tau$ is the relaxation time, $f_0 \sim 10^9$ is the attempt frequency, $\kB := 1.380649 \times 10^{-23}$ J/K is the Boltzmann constant and $T$ is the measurement temperature. The energy barrier $\Delta E = K_{\mathrm{eff}}(\mathbf{m}) V$ for different particle sizes is determined from Eq.~\ref{eq:effK} where $V$ is the particle volume. Fig.~\ref{fig:eb_shape} shows the aspect ratio dependence of the (a) azimuthal and (b) rotational volumetric energy barrier $\Delta E/V$ for elongated magnetite particles with shape anisotropy and for different particle shapes. The specific energy barrier for a given particle increases with the aspect ratio due to the formation of a uniaxial shape anisotropy and also due to an increase in the particle volume. Here we present the normalized energy barrier $\Delta E/V$ since the volume distribution of elongated particles is somewhat complicated to measure experimentally. The data in  Fig.~\ref{fig:eb_shape}(a) shows that even for small elongations the azimuthal energy barrier rapidly increases and dominates over the rotational barrier. As with the aspect ratio dependence of the demagnetizing factor there are significant differences in the energy barrier for different particle shapes, with the ellipsoidal particles showing the most rapid increase. The rotational energy barriers shown in Fig.~\ref{fig:eb_shape}(b) systematically decrease with increasing particle elongation due to the evolution of the energy minima towards the $\theta = 0, \pi$ directions with increasing shape anisotropy and are non-existent for elongations above 40\%. 

When considering magnetic hyperthermia applications it is essential to consider the measurement timescale since slow measurements conducted in a vibrating sample magnetometer will naturally show superparamagnetic effects that may not be evident at kHz frequencies used for generating heat~\cite{Burrows2010,Ruta2015}. The question arises whether small elongations of the particles are sufficient to significantly affect the intrinsic characteristic timescale of the magnetic relaxation. From the Arrhenius-N\'eel equation we can determine a characteristic timescale for a 12 nm diameter octahedral nanoparticle with an elongation of 20\% of $\tau \approx 16$ ns which is substantially below the characteristic timescale of hyperthermia measurements. Therefore it is likely that for heating applications with an oscillating magnetic field and high concentration the inter-particle dipole-dipole interactions~\cite{Burrows2010,Ruta2015,HaasePRB} are more important than the shape anisotropy of individual particles~\cite{Vallejo-FernandezAPL2013}. 

\subsection*{Spin dynamics of elongated nanocrystals}

\begin{figure*}[!h]
\center
\includegraphics[width=16.5cm, trim=0 0 0 0]{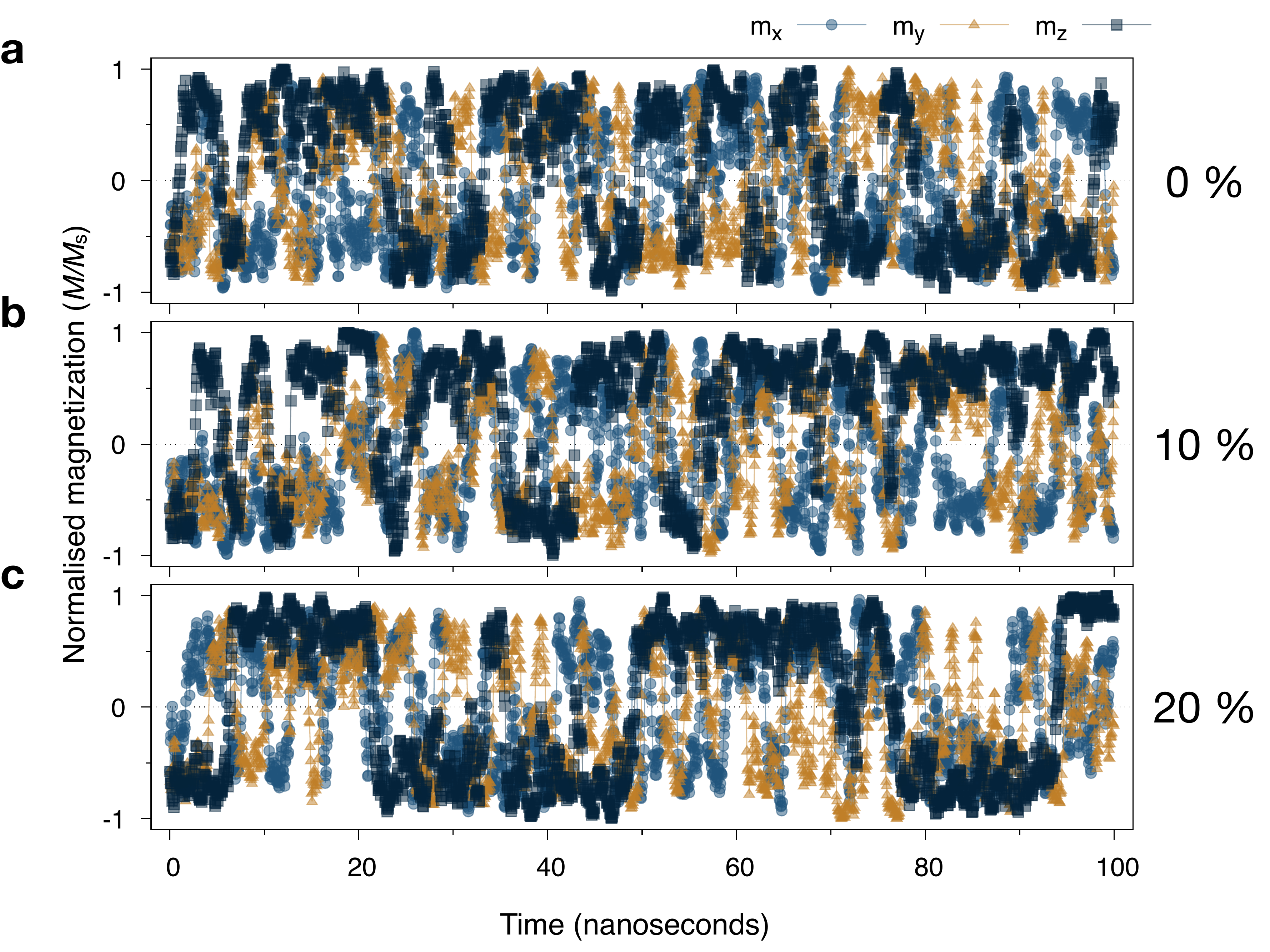}
\caption{Plot of the time dependence of the magnetization components for a 12 nm diameter octahedral nanoparticle for (a) 0\%, (b) 10\% and (c) 20\% elongation at $T = 300$ K. The $z$-component data show a transition from a fully superparamagnetic regime for the isotropic particle to a partially blocked regime for a particle with 20 \% elongation. The rotational components (transitions in x,y) are superparamagnetic in all cases.}
\label{fig:dynamics}
\end{figure*}

Finally we consider the dynamics of small \magnetite nanocrystals to investigate their thermal superparamagnetic behaviour and the transition to blocked magnetic behaviour due to elongation. To simulate the dynamics of magnetite nanoparticles we use the stochastic Landau-Lifshitz-Gilbert (sLLG) equation which describes the time-dependant behaviour of magnetic materials as described in the methods section. 

The switching dynamics of a 12 nm diameter (short axis) nanoparticle simulated over a period of 100 ns at a temperature of $T = 300$ K is shown in Fig.~\ref{fig:dynamics} for different particle elongations. The isotropic 12 nm particle in Fig.~\ref{fig:dynamics}(a) displays superparamagnetic behaviour in $,x,y$ and $z$ orientations of the magnetization due to the low magnetic anisotropy. However, over the 100 ns timescale of the simulation the noise shows a telegraph-like structure where the 8 easy-axis directions are metastable for 2-5 ns. The orientation of the magnetization away from the $\pm z$-axis is also clear with a mean value close to $m_z = \pm \sqrt{3}/3$ as expected from the cubic anisotropy.

The data for the 10\% elongated particle shown in Fig.~\ref{fig:dynamics}(b) shows fewer transitions along the $\pm z$-axis than the isotropic particle due to the additional magnetic shape anisotropy. An additional effect is also the increase of magnetic volume by 25\% due to the elongation, which also contributes to the effective magnetic anisotropy of the particle. When considering experimental samples it is important to disentangle these two effects where there are increased contributions to both volume and shape anisotropy.
% 45% larger V for 20%
For the largest elongation of 20\% shown in Fig.~\ref{fig:dynamics}(c), the particle shows significantly fewer transitions along the $\pm z$-axis and a much longer average metastable lifetime around $\tau \sim 14$ ns. In contrast the $x,y$ fluctuations of the magnetization are much larger and no longer exhibit telegraph noise, indicating a transition to fully superparamagnetic behaviour. The simulated relaxation time of $\tau \sim 14$ ns is close to the analytical value of $\tau \approx 16$ ns calculated from the Arrhenius-N\'eel equation. Interestingly this is significantly shorter than expected from experimental measurements of individual nanoparticles with typical lifetimes on the microsecond timescale~\cite{PiotrowskiIEEE2014}. This suggests that while shape anisotropy can be an important contribution to the magnetic anisotropy, it seems that surface anisotropy may still make a larger and more important contribution for nanoparticles with small elongations $<20$\% . Some samples do however exhibit large elongations averaging around 40\% with outliers with elongations of 100 \%, likely formed by aggregation of nuclei during the growth~\cite{Vallejo-FernandezAPL2013}. In such cases the relaxation time will extend to hundreds of nanoseconds to microseconds where the shape anisotropy will be the dominant contribution to the effective magnetic anisotropy of the particles. Magnetotactic bacteria typically show moderate particle elongations~\cite{ChariaouBio2015,AlphanderyFBB2014} where magnetic shape anisotropy makes a contribution to the stability but is not likely to be the dominant contribution.

\section*{Discussion}
In conclusion, we have developed an atomistic model of elongated magnetite nanocrystals with different shapes and surface faceting to investigate the role of shape anisotropy on the effective magnetic anisotropy energy. We have found that a cubic contribution to the effective anisotropy remains for relatively large elongations and that cannot be ignored when considering numerical and analytical models of magnetic hyperthermia. An approximate 4$^{\mathrm{th}}$-order uniaxial anisotropy has been derived to describe the largest energy barrier for elongated particles that retains a two state energy minimum compatible with simple numerical models. We also find that the shape anisotropy for faceted particles deviates significantly from Osborn's relation for an ellipsoid leading to an overestimation of the contribution to the effective anisotropy. In particular for moderately ($<20\%$) elongated nanocrystals at small sizes the additional shape anisotropy is insufficient to stabilize the particles even at short timescales, supporting the argument that inter-particle interactions or surface anisotropy make important contributions to N\'eel heating in magnetic hyperthermia applications. Previous experimental findings\cite{PiotrowskiIEEE2014} suggest that the relaxation time of similar sized and approximately spherical nanoparticles is significantly longer than 16 ns, meaning that our simplified model does not yet capture all the essential physics necessary to explain the relaxation time of actual magnetite nanoparticles. Additional surface structural relaxation and changes in the exchange coupling are therefore likely to play an important role in the determination of the effective anisotropy and the relaxation time.
Our findings are important for the future development of magnetic hyperthermia where nanoparticle properties should have reproducible heating properties with a narrow distribution of size and magnetic properties. In future work we will address the magnetic relaxation of magnetite nanoparticles considering size dependent energy barriers and the role of bulk defects~\cite{Nedelkoski2017} and surface effects~\cite{Oberdick2018} and larger particles with non-collinear magnetization structures where relaxation rates may be significantly different~\cite{GaraninPRBB2018}.

\section*{Methods}

\subsection*{Faceted particle generation algorithm}
\begin{figure}[!htb]
\center
\includegraphics[width=3.6cm, trim=0 0 0 0]{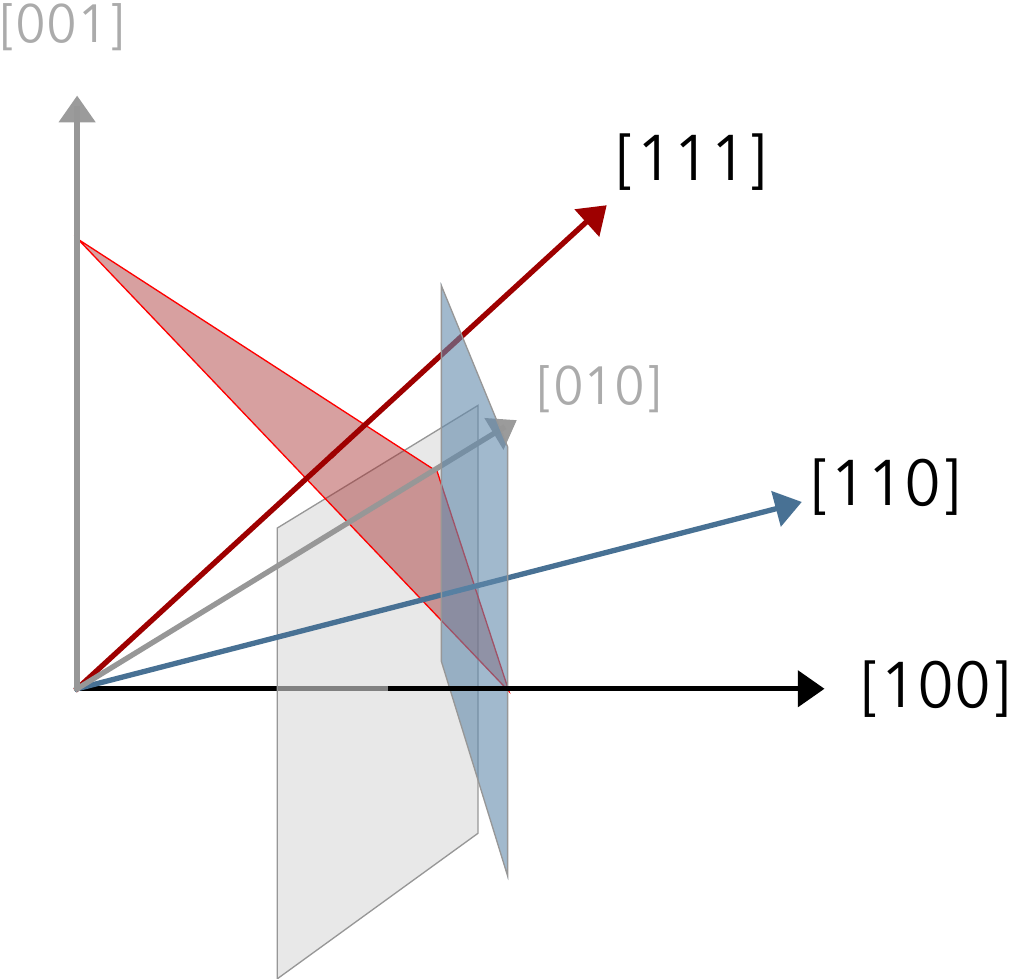}
\includegraphics[width=4.6cm, trim=0 0 0 0]{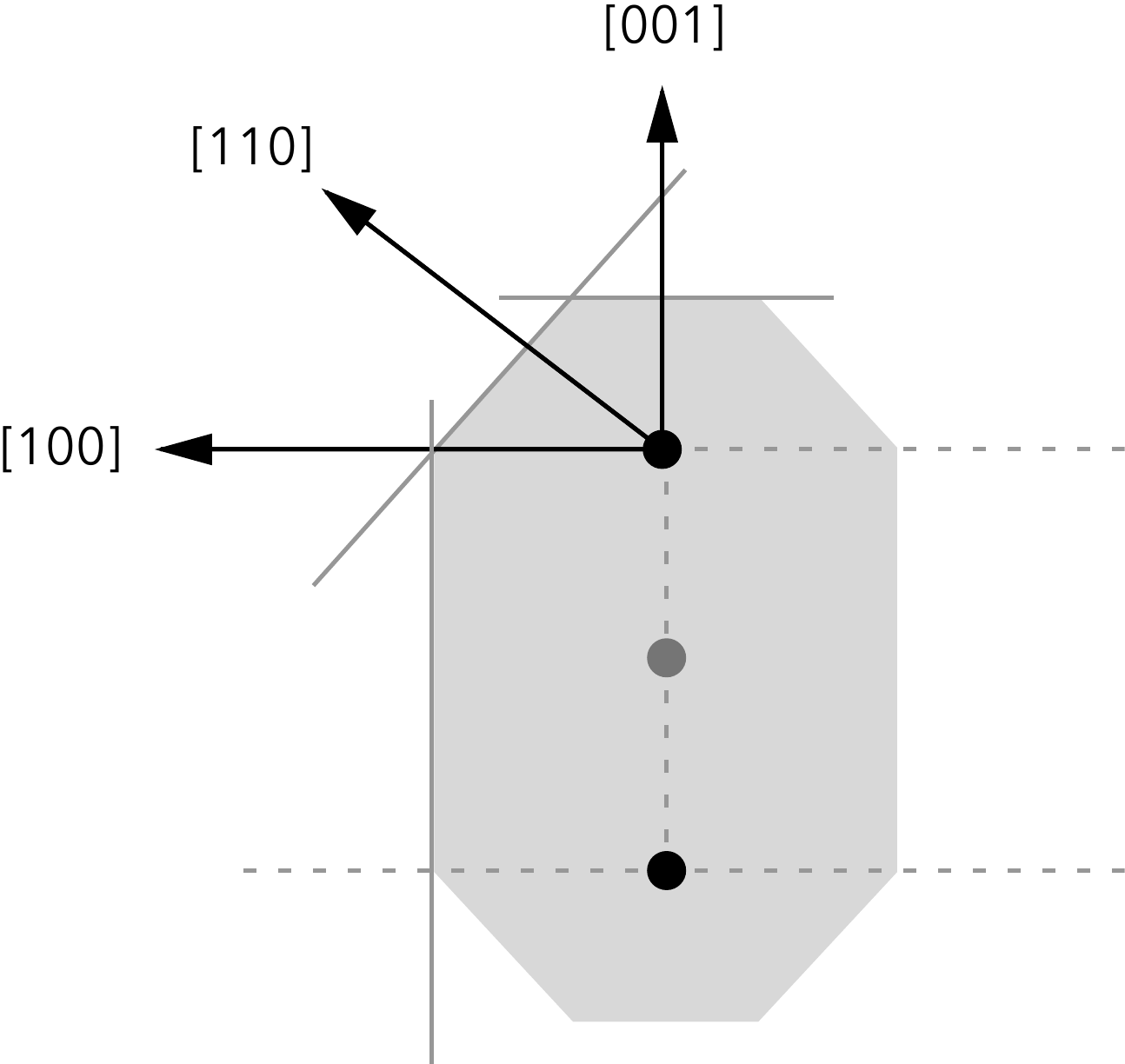}
\caption{Schematic illustration of how faceted nanoparticles are constructed. The particles are cut from a single crystal of magnetite. Planes perpendicular to the [100], [010], [001], [110] and [111] crystal directions define the limits of the particle along each direction as a function of particle size $r$ as shown in (a). Atoms between the origin and all defined planes are kept within the particle. The relative sizes $r_{100}$, $r_{110}$ and $r_{111}$ define the final shape of the crystal. Elongated nanoparticles are constructed in a similar way defining virtual shifted centres away from the origin (b), defining a long axis along the [001] crystal direction.}
\label{fig:algorithm}
\end{figure}

The method used to generate arbitrarily faceted nanocrystals is shown schematically in Fig.~\ref{fig:algorithm}(a). We first create a large single crystal of magnetite of size $4r$ along each spatial direction, where $2r$ is the intended particle size.  We then define fractional radii along the different crystal directions $r_{100}$, $r_{110}$ and $r_{111}$ defining the extent of truncating planes along the respective crystal directions in the positive quadrant ($|x|, |y|, |z|$). Atoms between these planes and the origin are retained within the particle and the rest are deleted, leaving a faceted nanoparticle. The choice of relative radii therefore selects the shape of the particle, for example if  $r_{100} = 2r$, $r_{110}= 2r$ and $r_{111}= r\sqrt 3$ this will construct a regular octahedron particle with diagonal diameter $2r$. Elongated particles are created in a similar way by defining a virtual origin along the $z$-axis of elongation, as shown in Fig.~\ref{fig:algorithm}(b). This retains the geometric faceting while allowing for arbitrary elongation of the particles.

\subsection*{Atomistic spin model}
 A Heisenberg spin Hamiltonian $\mathscr{H}$ describes the energetics of the system describing energy contributions \cite{EvansVMPR2013} for the octrahedrally (O) and tetrahedrally (T) coordinated iron sublattices:
%----------------------------
\begin{eqnarray}
\mathscr{H} &=& \mathscr{H}_{\mathrm{Fe}}^{\mathrm{T}} + \mathscr{H}_{\mathrm{Fe}}^{\mathrm{O}} \\
\nonumber
\mathscr{H}_{\mathrm{Fe}}^{\mathrm{T}} &=&  -\frac{1}{2}\sum_{i,j} \smJijFeTT \sms_i \cdot \sms_j - \frac{1}{2}\sum_{i,\delta} \smJijFeTO \sms_{i} \cdot \sms_{\delta}\\
&& +\kcc \sum_i \left(S_x^4 + S_y^4 +S_z^4\right) - \smmuFeT \sum_i \left( \smHapp + \smdip \right)\cdot \sms_i \\
\nonumber
\mathscr{H}_{\mathrm{Fe}}^{\mathrm{O}} &=& -\frac{1}{2}\sum_{\nu,\delta} \smJijFeOO \sms_{\nu} \cdot \sms_{\delta} - \frac{1}{2}\sum_{\nu,j} \smJijFeTO \sms_{\nu} \cdot \sms_{j}\\
 &&  +\kcc \sum_{\nu} \left(S_x^4 + S_y^4 +S_z^4\right) - \smmuFeO \sum_{\nu} \left( \smHapp + \smdip \right) \cdot \sms_{\nu}
\label{eq:hamiltonian}
\end{eqnarray}
%----------------------------
% 2019 parameters
%----------------------------
%material[1]:exchange-matrix[2] = -4.93e-21
%material[2]:exchange-matrix[2] = +0.986e-21
%material[2]:fourth-order-cubic-anisotropy-constant = -5.9206919e-25 
%----------------------------
where \sms are unit vectors describing the direction of the magnetic moments at each atomic site, $i,j$ label tetrahedral Fe sites with moment $\smmuFeT = 5~\muB$, $\nu,\delta$ label octahedral Fe sites with  moment \smmuFeO = 4.5~\muB, \smHapp is the externally applied magnetic field vector and $\smdip = \mu_0 \mathbf{B}_{\mathrm{dip}}$ is the local dipole field. $\smJijFeTO = -4.93$ zJ (zepto Joules) / link and $\smJijFeOO  = +0.986 $ zJ / link  are the Fe-Fe exchange interaction energies linking tetrahedral-octahedral and  octahedral-octahedral sites respectively. $\kc = -5.921 \times 10^{-25}$ J/atom describes the local cubic magnetocrystalline anisotropy at zero temperature with an easy axis along the (111) crystal directions. 

Atomistic spin dynamics are described by the sLLG equation applied at the atomistic level \cite{Ellis2015} and given by
\begin{equation}
\frac{\partial \Si}{\partial t} = -\frac{|\gamma|}{(1+\lambda^2)}[\Si \times \Heff + \lambda \Si \times (\Si \times \Heff)]
\label{eq:LLG}
\end{equation}
Here, \Si is a unit vector representing the direction of the magnetic spin moment of site $i$, $\gamma = 1.76 \times 10^{11}$ T$^{-1}$s$^{-1}$ is the gyromagnetic ratio and \Heff is the net magnetic field on each spin computed from the derivative of the full spin Hamiltonian in Eq.~\ref{eq:hamiltonian}. The first term of the sLLG describes the precession of the atomic spins and arises due to their quantum mechanical interaction with the effective magnetic field, and the second term describes direct angular momentum transfer between the spins and a heat bath which aligns the magnetization along the field direction. In this atomistic formulation of the sLLG, $\lambda = 0.063$ represents the local intrinsic contributions of spin-lattice and spin-electron interactions and is extracted from experimental data for the effective Gilbert damping for magnetite thin films \cite{LuAPL2019}. 

The effective field in the sLLG is given by
\begin{equation}
    \Heff = -\frac{1}{\mu_i}\frac{\partial \mathscr{H}}{\partial \Si} + \Hth\mathrm{.}
\end{equation}
where $\mu_i = \smmuFeO, \smmuFeT$ is the local atomic spin moment and \Hth is a stochastic thermal field simulating the effects of temperature using a Langevin dynamics formalism\cite{EvansVMPR2013}. To ensure the agreement of our classical simulations with experimental data for the temperature dependence of the saturation magnetization and anisotropy we apply spin temperature rescaling \cite{EvansPRB2015} to reduce the effective strength of the thermal spin fluctuations, replicating the underlying quantum nature of the spin system. The rescaled temperature $T_{\mathrm{sim}}$ is given by the expression 
\begin{equation}
T_{\mathrm{sim}} = \Tc \left(\frac{T}{\Tc}\right)^{\eta}
\end{equation}
where \Tc is the Curie temperature, $T$ is the temperature and $\eta = 2.00$ is the temperature rescaling exponent and assumed to be the same for both magnetic sublattices. The spin system is integrated using the Heun integration scheme\cite{garcia1998langevin} to solve the stochastic LLG and determine its time evolution. This scheme preserves the magnitude of the spins through renormalization and can be solved in a computationally efficient manner. To thermally equilibrate the system we utilize a Monte Carlo preconditioner using an adaptive move algorithm \cite{AlzateCardonaJPCM2019}. The calculations have been carried out using the \vampire software package \cite{vampire-url,EvansVMPR2013}.

\subsection*{Inter \& intra macro-cell method for dipole field calculations}
In magnetite the magnetic anisotropy energy is low meaning that small nanoparticles are usually superparamagnetic on typical experimental measurement timescales of seconds. Uniaxial magnetic shape anisotropy in elongated nanoparticles is therefore often assumed to be a dominant energy contribution when modelling fine particle systems. In order to accurately assess the contribution of the shape anisotropy to the stability of magnetite nanoparticles we have implemented the intra and inter-macrocell algorithm developed by Bowden \textit{et al} \cite{Bowden2016} to calculate the effective shape anisotropy of elongated magnetic nanoparticles. The method calculates the dipole tensor between neighbouring macro-cells with atomic scale accuracy and gives an exact representation of the dipole field provided that the atomic magnetic moments within each macro-cell are in perfect alignment. The domain wall width of magnetite is in excess of 50 nm and so over a small volume such as a nanoparticle the magnetization can be considered as being uniform. In the inter \& intra macro-cell approach the system is discretized into small macro-cells with a size much less than the domain wall width. A dipole tensor describes the interactions within and between macro—cells and retains the atomic information in the form of real-space coordinates. The contribution to the dipole-dipole interaction is separated into two parts: $a$) one that arises from the interaction of the atomic moments within a macro-cell with the atomic moments in another cell, called inter macro-cell  contribution, and $b$) one determined by the interaction among spins within the same macro-cell, defined as the intra macro-cell contribution. 
Following this method, we can write the dipolar tensor for a macro-cell $p$ as summation of the contribution from interactions with magnetic moments in the other macro-cells $q$ (inter) and inside the macro-cell $p$ (intra). 
In terms of the effective magnetic field terms $\mathbf{B}$, we can write:
\begin{eqnarray}
\Bdipcellvec^p &= \Bdipvec^q\left(\mathrm{inter}\right) + \Bdipvec^p\left(\mathrm{intra}\right) \nonumber
\\
%\nonumber \\ 
&= \Diptens_{qp}^{\mathrm{inter}}\cdot \mcellvec^q + \Diptens_{pp}^{\mathrm{intra}}\cdot \mcellvec^p
\label{eq:dip}
\end{eqnarray}
where $\Diptens_{qp}^{\mathrm{inter}}$ and $\Diptens_{pp}^{\mathrm{intra}}$ are effective dipole-dipole matrices, given by:
\begin{align}
\Diptens_{qp}^{\mathrm{inter}} &= \frac{1}{n_p n_q}\sum_{qj=1}^{n_q} \sum_{pi=1}^{n_p} \Diptens_{qj,pi}^{\mathrm{inter}}
\\
%\nonumber \\
\Diptens_{pp}^{\mathrm{intra}} &= \frac{1}{n_p n_q}\sum_{pj\ne pi}^{n_p} \sum_{pi=1}^{n_p} \Diptens_{pj,pi}^{\mathrm{intra}} \,.
\end{align}
Here $1\leqslant pi\,,pj\leqslant n_p$, $1 \leqslant qj \leqslant n_q$ are the indexes running over the individual atoms within the cells $p$ and $q$ which enclose $n_p$ and $n_q$ spins, respectively.
$\Diptens_{qj,pi}^{\mathrm{inter}}$ and $\Diptens_{qj,pi}^{\mathrm{intra}}$ are dipole-dipole matrices calculated on the real-space atoms coordinates for interactions with other cells and within the same macro-cell, respectively. 
The dipolar matrix for the interaction between different macro-cells, is given by:
\begin{equation}
\label{eq:dipole_tensor_inter}
\Diptens_{qj,pi}^{\mathrm{inter}} = \frac{1}{4\pi r^3_{piqj}} 
\left(
	\begin{array}{ccc} 
	3x^2 - 1    & 3x y  & 3x z\\ 
	3x y & 3y^2 - 1     & 3y z\\
    3x z & 3y z  & 3z^2 - 1
	\end{array}
\right),
\end{equation}
where $\vec{r}_{piqj}$ is the distance between the individual atomic dipole moments at positions $\mathbf{r}_{pi}$ and $\mathbf{r}_{qj}$ and $(x,y,z)$ 
are the unitarian Cartesian components of $\mathbf{r}_{qj}$.
To obtain the intra-macro-cell dipole matrix one just replaces $qj$ with $pj$.
In Equation~\ref{eq:dipole_tensor_inter} we have used the symmetry of the dipole tensor components according to where the off-diagonal terms are symmetric: $xy=yx$, $xz=zx$ and $yz=zy$. This allows the independent components of the dipolar tensor to reduce from nine to six to reduce the amount of data to be stored. We note that the dipole-dipole term differs from the Maxwellian continuum model in the actual value of the demagnetizing field and demagnetizing energy due to the absence of the self-demagnetization term. Our approach works independently of the shape of the macro-cell and the position of the centre of the cell does not affect the calculation because of how the dipole-dipole matrix is calculated. It is worth stressing that the inter and intra macro-cell approach leads to agreement with the dipole-dipole interaction on the atomic scale in case of uniform magnetization in the macro-cells\cite{Bowden2016}. Also, it is important to observe that in order to match the criterion of uniform magnetization within a cell, in real systems the macro-cell size should be chosen smaller than the domain wall width. As noted by~\cite{Bowden2016}, both the bare macro-cell model and the inter and intra macro-cell approach yield the same dipolar field for macro-cells whose distance is more than twelve in macro-cell units. 
This can be exploited in order to simplify the initialization of the dipole tensors. We can define a cut-off range within which the inter and intra macro-cell method is used, whereas for macro-cells that are outside this range the inter macro-cell term of the dipole-dipole interaction can be replaced by a direct dipole-dipole macro-cell calculation. With this approach, thanks to the intra-macro-cell term in particular, it is possible to achieve an accurate description of the dipolar field and dipolar energy for surfaces and irregular shaped regions essential for the calculation of the dipole field in nanoparticle systems.

\section*{Acknowledgments}
The financial support of the Engineering and Physical Sciences Research Council (Grant No. EPSRC EP/P022006/1) is gratefully acknowledged. Daniel Meilak gratefully acknowledges the financial support of EPSRC DTP Studentship (EP/N509802/1) project 1800054. SAM acknowledges support from US Department of Energy grant \# DE-FG02-08ER46481. This work was performed using the \textsc{viking} supercomputing cluster provided by the University of York and enabled by code enhancements implemented and funded under the ARCHER embedded CSE programme (eCSE0709 and eCSE1307).  The authors would like to thank Kevin O'Grady for helpful discussions.

\section*{Competing interests}
The author(s) declare no competing interests.

\section*{Author contributions}
RMO, SJP, SJ, DM and RFLE conducted the investigation, performed the simulations and generated the data. RMO, DM and RFLE developed the atomistic magnetite model and parameters. SJP developed the code for calculating the energy barriers for the particles. AM and SJ implemented the inter and intra macrocell method for the dipole-dipole field calculation. RFLE wrote the paper and all authors contributed to the manuscript, analysis and interpretation of results.

%\clearpage
%\bibliography{/Users/rfle500/Documents/Work/Papers/Bibliography/library}
\bibliography{library,local}
%---------------------------------------------------------------------------%

%\begin{thebibliography}{11}%
%\end{thebibliography}%

\end{document}